\documentclass{aa}
\usepackage{txfonts}
\usepackage{psfig}
\usepackage{epsfig}

\def\Swift{\emph{Swift}}
\def\etal{et al.\ }
\def\til{\ensuremath{\sim\,}}
\def\sqiglt{\hbox{\rlap{\lower.55ex \hbox {$\sim$}}\kern-.05em \raise.4ex \hbox{$<$}\,}}
\def\sqiggt{\hbox{\rlap{\lower.55ex \hbox {$\sim$}}\kern-.05em \raise.4ex \hbox{$>$}\,}}
\def\cps{counts s$^{-1}$}
\newcommand{\tim}[1]{\ensuremath{\times 10^{#1}}}

\begin{document}

\title{An online repository of \Swift/XRT light curves of GRBs.}

\author{P.A. Evans\inst{1}\thanks{pae9@star.le.ac.uk} \and A.P.
Beardmore\inst{1} \and K.L. Page\inst{1} \and L.G. Tyler\inst{1}\and J.P.
Osborne\inst{1} \and M.R. Goad\inst{1} \and P.T. O'Brien\inst{1} \and L. Vetere\inst{2} \and J.
Racusin\inst{2} \and D. Morris\inst{2} \and D.N. Burrows\inst{2} \and M.
Capalbi\inst{3} \and M. Perri\inst{3} \and N. Gehrels\inst{4} \and P.
Romano\inst{5,6}}

\institute{Department of Physics and Astronomy, University of Leicester,
Leicester, LE1 7RH, UK \and Department of Astronomy and Astrophysics, 525 Davey
Lab., Pennsylvania State University, University Park, PA 16802, USA \and 
ASI Science Data Center, ASDC c/o ESRIN, via G. Galilei, 00044 Frascati, Italy \and
NASA/Goddard Space Flight Center, Greenbelt, MD 20771, USA \and INAF-Osservatorio Astronomico di
Brera, via E. Bianchi 46, 23807 Merate (LC), Italy \and Universit\`a degli Studi di Milano, Bicocca,
Piazza delle Scienze 3, I-20126, Milano, Italy}

\date{Received / Accepted}

\abstract
%Context
{\Swift\ data are revolutionising our understanding of Gamma Ray Bursts. Since
bursts fade rapidly, it is desirable to create and disseminate accurate
light curves rapidly.}
%AIMS
{To provide the community with an online repository of X-ray light curves
obtained with \Swift. The light curves should be of the quality expected of
published data, but automatically created and updated so as to be
self-consistent and rapidly available.}
%METHODS
{We have produced a suite of programs which automatically generates \Swift/XRT
light curves of GRBs. Effects of the damage to the CCD, automatic
readout-mode switching and pile-up are appropriately handled, and the data are
binned with variable bin durations, as necessary for a fading source. }
%RESULTS
{The light curve repository website\thanks{http://www.swift.ac.uk/xrt\_curves}
contains light curves, hardness ratios and deep images for every GRB which
\Swift's XRT has observed. When new GRBs are detected, light curves are created
and updated within minutes of the data arriving at the UK Swift Science Data
Centre.}
{}


\maketitle

\section{Introduction}
\label{sec:intro}

\begin{figure}
\resizebox{\hsize}{!}{\includegraphics[angle=-90,width=8.0cm]{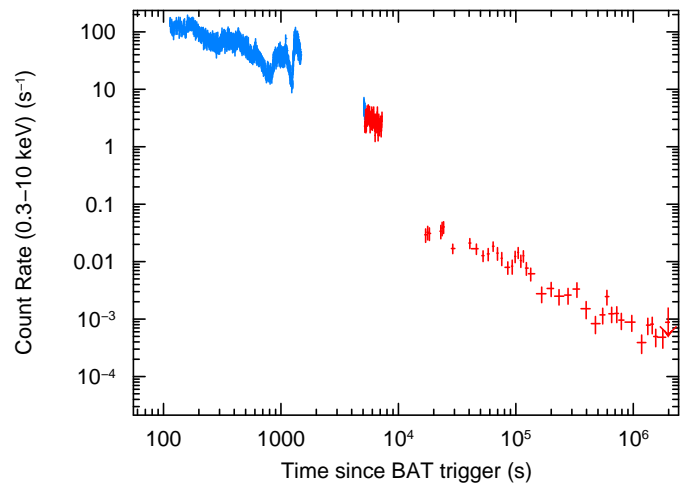}}

\caption{\Swift\ X-ray light curve of GRB 051117a (Goad \etal2007), created using
the software described in this paper and obtained from the \Swift\ Light Curve
Repository.}

\label{fig:GRB051117}
\end{figure}

The data from the \Swift\ satellite (Gehrels \etal2004), and particularly its
X-ray Telescope (XRT, Burrows \etal2005), are revolutionising our understanding
of Gamma Ray Bursts (GRBs, see Zhang 2007 for a recent review). The XRT typically
begins observing a GRB \til 100 s after the trigger, and usually follows it for
several days, and occasionally for months (e.g., Grupe \etal2007). However,
creating light curves of the XRT data is a non-trivial process with many
pitfalls. The UK Swift Science Data Centre is automatically generating
light curves of GRBs -- an example light curve is given in
Fig.~\ref{fig:GRB051117} -- and making them immediately available online. In
this paper we detail how the light curves are created, and particularly, how the
complications specific to these data are treated.

\subsection{Aspects of light curve generation}
\label{sec:complicate}

In general, creation of X-ray light curves is a relatively simple, quick task
using {\sc ftools} such as the {\sc xselect} and {\sc lcmath} packages. Building
\Swift/XRT light curves of GRBs, however, has a number of complications which can
make the task difficult and slower, as described below. 

\subsubsection{GRBs fade}
\label{sec:fade}

The standard light curve tools, such as those mentioned above, produce light
curves with uniform bin durations. Since GRBs fade by many orders of magnitude,
long-duration bins are needed at late times in order to detect the source.
However, GRBs show rapid variability and evolution at early times, and short
time bins are needed to resolve these features. A better approach to producing
GRB light curves is to bin data based on the number of counts in a bin, rather
than the bin duration. This is common practice for X-ray spectroscopy, however
there are no {\sc ftools} available to do this for light curves. While
this is our chosen means of binning GRB light curves, it is not the only option.
For example, one could use the Bayesian blocks method (Scargle 1998) to
determine the bin size.

\begin{figure}
\resizebox{\hsize}{!}{\includegraphics[width=8.0cm]{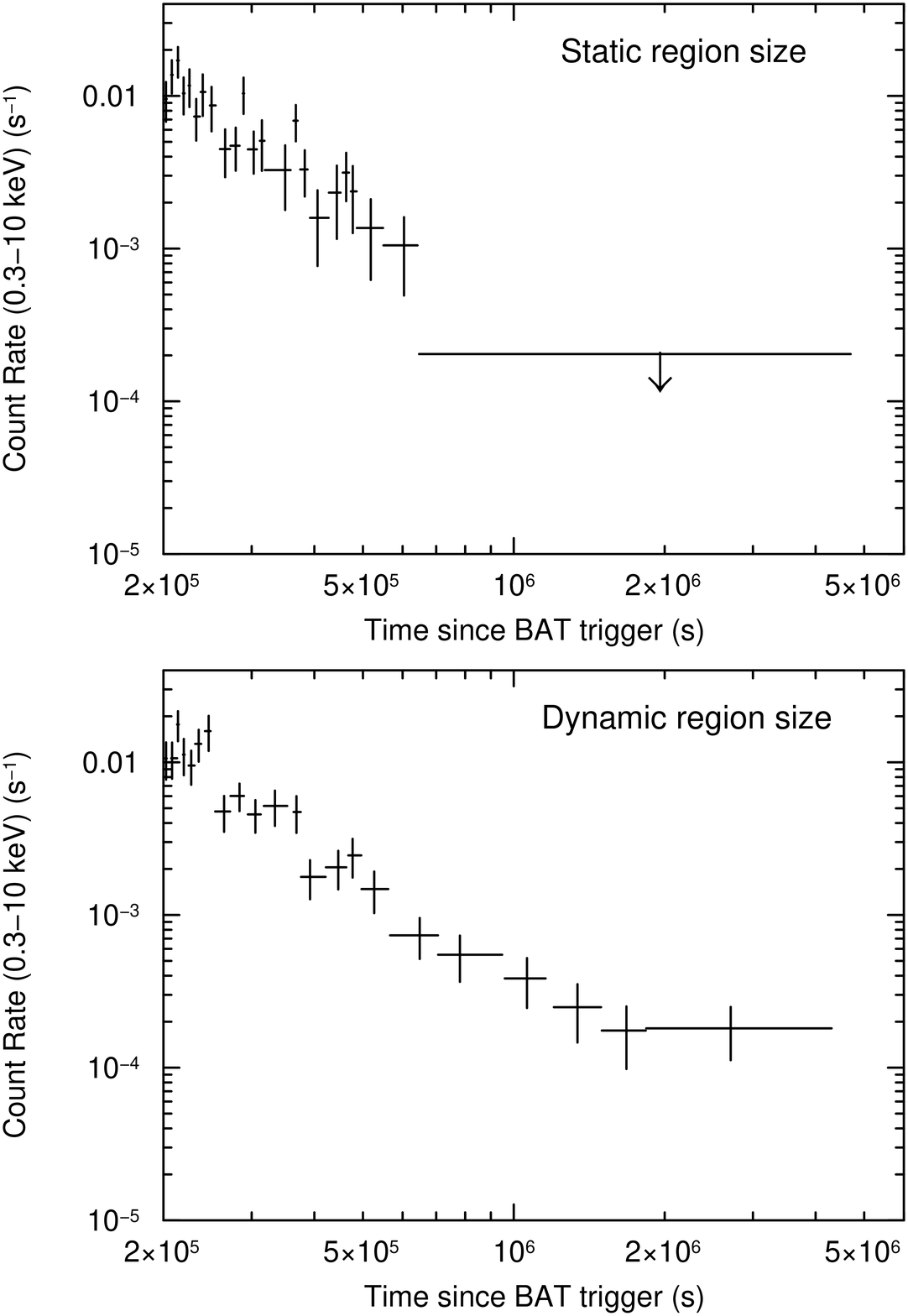}}
%\resizebox{\hsize}{!}{\includegraphics[angle=-90,width=8.0cm]{fig2b.eps}}
\caption{Late-time \Swift\ X-ray light curves of GRB 060614 (Mangano \etal2007), showing the
need for the source region to be reduced as the data fades.
\newline \emph{Top panel}: Where the source extraction region remains large at
late times, the source cannot be detected after 600 ks.
\newline \emph{Bottom panel}: Using a smaller source extraction region at later
times suppresses the background, yielding 6 more datapoints on the light curve.}
\label{fig:sourcerad}
\end{figure}

Another complication caused by the fading nature of GRBs is that when the burst
is bright, it is best to extract data for a relatively large radius around the
GRB position, to maximise the number of counts measured. When the GRB has faded,
using such a large region means that the measured counts would be dominated by
background counts, making it harder to detect the source, thus it is necessary
to reduce the source region size as the GRB fades. This is illustrated in
Fig.~\ref{fig:sourcerad}.

\subsubsection{\Swift\ data contain multiple observations and snapshots}
\label{sec:observe}

The \Swift\ observing schedule is planned on a daily basis, and each day's
observation of a given target has its own observation identification (ObsID) and event list.
Thus if \Swift\ follows a GRB for two weeks, it will produce up to fourteen
event lists, all of which need to be used in light curve creation. At late times
it may become necessary to combine several datasets just to detect the GRB. 

Also, \Swift's low-Earth orbit means that it is unable to observe most targets
continuously. Thus, any given ObsID may contain multiple visits to the target
(`snapshots') which again will need to be combined (this differentiation between
observations -- datasets with a unique ObsID -- and snapshots -- different
on-target times within an ObsID -- will be used throughout this paper).  Combining
snapshots/observations can result in bins on a light curve where the fractional
exposure is less than 1. This must be taken into account in calculating the
count rate. 

The standard pipeline processing of \Swift\ 
data\footnote{http://swift.gsfc.nasa.gov/docs/swift/analysis/xrt\_swguide\_v1\_2.pdf}
ensures that the sky coordinates are correctly attained for each event, however
the position of the GRB on the physical detector can be different each snapshot
due to changes in the spacecraft attitude. This becomes a problem when one
considers the effects of bad pixels and columns.

\subsubsection{CCD Damage}
\label{sec:damage}

On 2005-May-27 the XRT was struck by a micrometeoroid (Abbey \etal2005).
Several  of the detector columns became flooded with charge (`hot'), and have
had to be permanently screened out. Unfortunately, these lie near the centre of
the CCD, so the point spread function (PSF) of a GRB often extends over these
bad columns. As well as these columns there are individual `hot pixels' which
are screened out, and other pixels which become hot when the CCD temperature
rises, so may be screened out in one event list, but not in the next. Exposure
maps and the {\sc xrtmkarf} tool can be used to correct for this, however this
has to be done individually for each \Swift\ snapshot (since the source will not
be at the same detector position from one snapshot to the next). A single day's
observation contains up to 15 snapshots, thus to do this manually is a slow,
laborious task. The forthcoming {\sc xrtlccorr} program should make this process
easier, however it will still need to be executed for each observation.

\subsubsection{Automatic readout-mode switching}
\label{sec:modes}

One of the XRT's innovative features is that it changes readout mode
automatically depending on the source intensity (Hill \etal2004). At high
count-rates it operates in Windowed Timing (WT) mode, where some spatial
information is sacrificed to gain time resolution ($\Delta t=1.8\tim{-3}$ s). At
lower count-rates Photon Counting (PC) mode is used, yielding full spatial
information, but lower time resolution ($\Delta t=2.5$ s). The XRT also has
Photodiode (PD) mode, which contains no spatial information, but has very high
time resolution ($\Delta t=1.4\tim{-4}$ s). This mode was designed to operate
for higher count-rates than WT mode, however it was disabled following the
micrometeoroid impact. Prior to this, the XRT produced very few PD mode frames
before switching to WT so we have limited our software to WT and PC modes.

\begin{figure}
\resizebox{\hsize}{!}{\includegraphics[width=8.0cm]{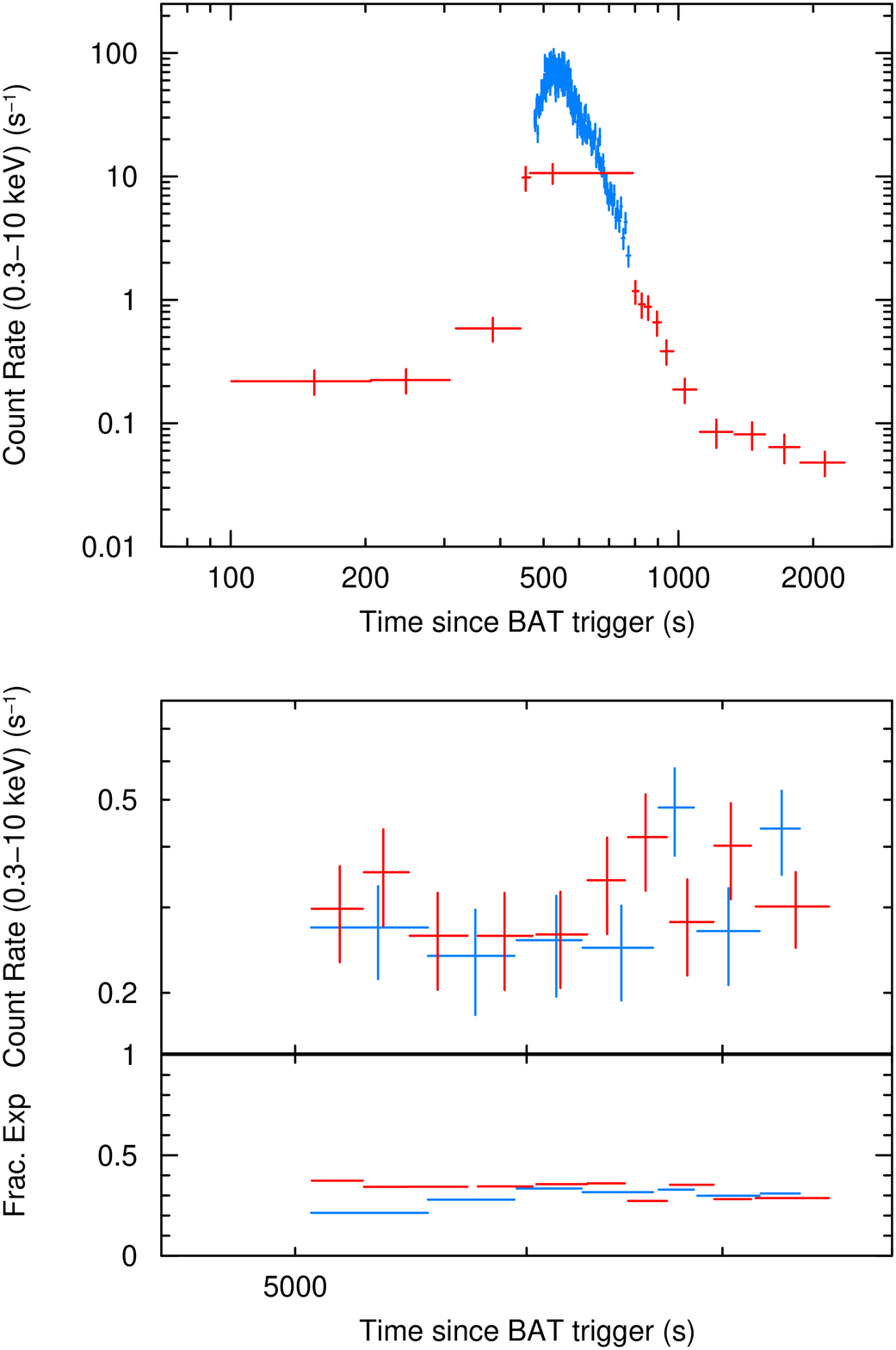}}
%\resizebox{\hsize}{!}{\includegraphics[angle=-90,width=8.0cm]{fig3b.eps}}
\caption{\Swift\ X-ray light curves of two GRBs, showing the switching between
readout modes. WT mode is blue, PC mode red. 
\newline \emph{Top panel}: GRB 060929. The XRT changed from PC to WT mode due to
a large flare.
\newline \emph{Bottom panel}: GRB 050315. The XRT was `mode-switching' during
the second snapshot. The lower pane shows the fractional exposure, which is
highly variable due to this effect.}
\label{fig:modeswitch}
\end{figure}

For a simple, decaying GRB the earliest data are in WT mode and as the burst
fades the XRT switches to PC mode. This is not always the case; the XRT
can toggle between modes. GRB~060929 for example, had a count-rate of \til0.1
\cps\, and the XRT was in PC mode, when a giant flare pushed the count-rate up
to \til100 \cps\ and the XRT switched into WT mode, causing a \til 200 s gap in
the PC exposure (Fig.~\ref{fig:modeswitch}, upper panel). Since the initial CCD
frames are taken in WT mode, and the PC data both preceded and succeeded the WT
data taken during the flare, there are large overlaps between the WT and PC
data. 

On occasions, such as when \Swift\ was observing GRB 050315 (Vaughan \etal2006),
the XRT  oscillates rapidly between WT and PC modes (Fig.~\ref{fig:modeswitch},
lower panel). This `mode switching' occurs when the count-rate in the central
window of the CCD changes rapidly. Such variation is  usually due to the rapid
appearance and disappearance of hot pixels at high ($\sim-52^\circ $C) CCD
temperatures (the XRT is only passively cooled due to the failure of the
on-board thermoelectric cooler, Kennea \etal2005), although contamination by
photons from the illuminated face of the Earth can also induce mode switching.
Recent changes in the on-board calibration have significantly reduced the effects
of hot-pixel induced mode switching, however when it does happen it complicates
light curve production by causing a variable fractional exposure. Also, during
mode switching the XRT does not stay in either mode long enough to collect
sufficient data to produce a light curve bin (see Section~\ref{sec:produce}),
thus the WT and PC bins can overlap. The lower panel of
Fig.~\ref{fig:modeswitch} illustrates these points.

\subsubsection{Pile-up}
\label{sec:pileup}

Pile-up occurs when two photons are incident upon on the same or adjacent CCD
pixels in the same CCD frame. Thus, when the detector is read out the two
photons are recorded as one event. Pile-up in the \Swift\ XRT has been discussed
by Romano \etal(2006) for WT mode and Vaughan \etal(2006) and Pagani \etal(2006)
for PC mode. Their quantitative analyses show the effects of pile-up at
different count rates, and we used these values to determine when we consider
pile-up to be a problem (see Section~\ref{sec:produce}). 

This problem is not unique to \Swift, but because GRBs vary by many orders of
magnitude, pile-up must be identified and corrected in a time-resolved manner. 
The standard way to correct for pile-up is to use an annular source extraction
region, discarding the data near the centre of the PSF where pile-up occurs. 
For constant sources, or those which vary about some roughly constant mean, it is
usually safe to use this annular region at all times. This is not true for GRBs,
which can span five decades in brightness; using an annulus when the burst is
faint would make it almost undetectable!

\vspace{2cm}

In the following sections we detail the algorithm used to generate light curves
automatically, and in particular we concentrate on how the above issues are
resolved.

\section{Light curve creation procedure}
\label{sec:creation}

The raw \Swift/XRT data are processed at the Swift Data Center at NASA's
Goddard Space Flight Center, using the standard \Swift\ software developed at
the ASI Science Data Center (ASDC) in Italy. The processed data are then sent to
the \Swift\  quick-look archives at Goddard, the ASDC, and the UK. As soon as
data for a new GRB arrive at the UK site, the light curve generation software is
triggered, and  light curves made available within minutes.

The light curve creation procedure can be broken down into three phases. The
\emph{preparation phase\/} gathers together all of the observations of the GRB,
creating summed source and background event lists. The \emph{production phase\/}
converts these data into time-binned ASCII files, applying corrections for the
above-mentioned problems in the process. The \emph{presentation phase\/} then
produces light curves from the ASCII files, and transfers them to the online
light curve repository.

\subsection{Phase \#1 -- Preparation Phase}
\label{sec:prepare}

In overview: this phase collates all of the observations, defines appropriate
source and background regions (accounting for pileup where necessary), and
ultimately produces a source event list and background event list for WT and PC
mode, which are then passed to the production phase.

The preparation phase begins by creating a list of ObsIDs for the GRB, and then
searching the file metadata to ascertain the position of the burst, the trigger
time, and the name. An image is then created from the first PC-mode event list
and the {\sc ftool xrtcentroid} is used to obtain a more accurate position. A
circular source region is then defined, centred on this position, and initially
30 pixels (71") in radius. A background region is also defined, as an annulus
centred on the burst with an inner radius of 60 pixels (142") and an outer
radius of 110 pixels (260"). The image is also searched for serendipitous
sources close to the GRB (e.g.\ there is a flare star 40" from GRB 051117A; Goad
\etal2007), and if any are found to encroach on the source region, the source
extraction radius is reduced to prevent contamination.

The software then takes each event list in turn. The bad pixel information is
obtained from the `BADPIX' FITS extension and stored for use in the production
phase. An image of the background region is created, and the {\sc detect}
routine in {\sc ximage} is used to identify any sources with a count-rate
$\geq3\sigma$ above the background level. For each source thus found, a circular
region centred on the source with a radius equal to the source extent returned
by {\sc ximage}, is excluded from the background region.

The event list is then broken up into individual snapshots, the mean count-rate
during each snapshot is ascertained and used to determine an appropriate
source region size (Table~\ref{tab:sreg}) and appended to the event list for
later use. The values in Table~1 were determined by manual analysis of many GRB
observations, and reflect a compromise between minimising the background level
while maximising the proportion of source counts that are detected.

For each snapshot, the detector coordinates of the object are found using the
{\sc pointxform ftool}, and used to confirm that both source and background
regions lie within the CCD. If the background region falls off the edge of the
detector it is simply shifted by an appropriate amount (ensuring that the inner
ring of the annulus remains centred on the source). The source region must
remain centred on the source in order for later count-rate corrections to be
valid, however if this results in part of the extraction region lying outside of
the exposed CCD area, the source region for this snapshot is reduced.

\begin{figure}
\resizebox{\hsize}{!}{\includegraphics[angle=-90,width=8.0cm]{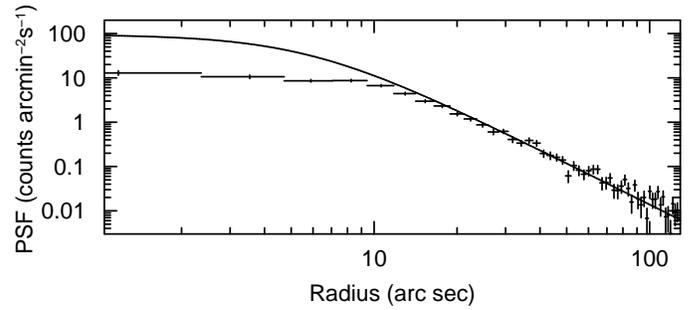}}
\caption{The PSF of GRB 061121 (Page \etal2007), during the first snapshot of PC
data. The model PSF was fitted to the data more than 25" from the burst. The
central 10" are clearly piled up.}
\label{fig:pileup}
\end{figure}

The first part of pile-up correction is carried out at this stage. A simple,
uniform time-bin light curve is created with bins of 1~s  (5~s) for WT (PC)
mode, and then parsed to identify times where the count rate climbs above 150
(0.6) \cps; such times are considered to be at risk of pile-up. For WT mode we
are unable to investigate further, since we have only one-dimensional spatial
information arranged at an arbitrary (albeit known) angle in a two dimensional
plane, and no tools currently exist to extract a PSF from such data. Instead,
the centre of the source region is excluded, such that the count rate  in the
remaining pixels never rises above 150 \cps. The number of excluded pixels
is typically in the range \til6--20, depending on the source brightness. For
PC mode, a PSF profile is obtained for the times of interest, and the wings of
this (from 25" outwards) are fitted with a King function which accurately
reproduces \Swift's PSF (Moretti \etal2005). This fit is then extrapolated back
to the PSF core, and if the model exceeds the data by more than the 1-$\sigma$
error on the data, the source is classified as piled up (Fig.~\ref{fig:pileup}).
The source region is then replaced with an annular region whose inner radius is
that at which the model PSF and the data agree to within 1-$\sigma$ of the data.
Note that these annular regions are only used during the intervals for which
pile-up was detected, the rest of the time a circular region is used (or a
box-shaped region for WT mode). If there are several separate intervals of
pile-up (e.g., pile-up lasts for several snapshots, or a flare causes the
count-rate to rise into the pile-up r\'egime), they each have their own annular
region. The inner radii of the annuli (or size of the excluded region in WT
mode) are stored in the event list, so that in the production phase the
count-rate can be corrected for events lost by the exclusion of the central part
of the PSF.

The time-dependent region files thus created are used to generate source and
background event lists for this snapshot. This process is performed for every
snapshot in every observation of the GRB, and the event lists are then combined
to yield one source and one background event list for each XRT mode.
Additionally, all PC-mode event lists are merged for use in the presentation
phase (phase \#3)

\begin{table}
\caption{Source extraction radii used for given count rates. $R$ is the
measured, uncorrected count rate.}
\label{tab:sreg}
\centering
\begin{tabular}{c c}
\hline\hline
Count rate $R$ (\cps)  &  Source radius in pixels (arc sec) \\
\hline
$R > 0.5$           & 30 (70.8") \\
$0.1 < R \leq 0.5$  & 25 (59.0")\\
$0.05 < R \leq 0.1$  & 20 (47.2")\\
$0.01 < R \leq 0.05$  & 15 (35.4")\\
$0.005 < R \leq 0.01$  & 12 (28.3")\\
$0.001 < R \leq 0.005$  & 9 (21.2")\\
$0.0005 < R \leq 0.001$  & 7 (16.5")\\
$ R \leq 0.0005$  & 5 (11.8")\\
\hline
\end{tabular}
\end{table}

\subsection{Phase \#2 -- Production Phase}
\label{sec:produce}

In this phase the data are first filtered so that only events with energy in the
range 0.3--10 keV are included. For WT (PC) mode, only events with grades 0--2
(0--12) are accepted. Each mode is then processed separately: WT and PC mode
data are not merged. The process described in this section occurs three times in
parallel: once on the entire dataset, once binning only soft photons (with
energies in the range 0.3--1.5 keV), and once binning only the hard photons
(1.5--10 keV). The data are then binned and background subtracted. Since the
source region is dynamic and could change within a bin, each background photon
is individually scaled to the source area (the source radius used was saved in
each event list during the preparation phase).

A bin (i.e.\ a point on the light curve) is defined as the smallest possible
collection of events which satisfies the following criteria:

\begin{itemize}
\item{There must be at least $C$ counts from the source event list.}
\item{The bin must span at least 0.5 (2.51) s in WT (PC) mode.}
\item{The source must be detected at a significance of at least 3$\sigma$.}
\item{There must be no more events within the source region in this CCD frame}
\end{itemize}

For the energy-resolved data, both the soft and hard data must meet these
criteria individually to complete a bin.

$C$, the minimum number of counts in the source region, is a dynamic parameter.
Its default value of 30 for WT mode and 20 for PC mode is valid when the source
count-rate is one count per second. It scales with count rate, such that an
order of magnitude change in count rate produces a factor of 1.5 change in $C$.
This is done discretely, i.e. where $1\leq {\rm rate} <10$, $C$=30 counts (WT
mode), for $10\leq {\rm rate} <100$, $C=45$ counts etc. $C$ must always be above
15 counts, so that Gaussian statistics remain valid. Note that $C$ always refers
to the number of measured counts, with no corrections applied, however `rate'
refers to the corrected count rate (see below). These values of $C$ give poor
signal-to-noise levels in the hardness ratio, so for the energy-resolved data we
require $2C$ counts in each band in order to create a bin.

The second criterion (the bin duration) is in place to enable reasonable
sampling of the background. For the third criterion we define the detection
significance as $\sigma=N/\sqrt{B}$, where $N$ is the number of net counts from
the source and $B$ is the number of background counts scaled to the source
area. Thus we require that a datapoint have a $<0.3\%$ probability of being a
background fluctuation before we regard it as `real'.

The final criterion is used because the CCD is read out at discrete times, thus
all events that occur between successive read-outs (i.e.\ within the same frame)
have the same time stamp. Thus, if the final event in one bin and the first
event in the next were from the same frame, those bins would overlap. Apart from
being cosmetically unpleasant, this will also make modelling the light curves
much harder, and is thus avoided.

At the end of a \Swift\ snapshot, there may be events left over which do
not yet comprise a full bin. These will be appended to the last full bin from
this snapshot, if there is one, otherwise they are carried over to the next
snapshot. At the end of the event list, if there are still spare events, this
bin is replaced with an upper limit on the count rate. This is calculated at the
3$\sigma$ (i.e. 99.7\%) confidence level, using the Bayesian method championed by
Kraft, Burrows and Nousek (1991). 

As the data are binned and background subtracted, the count-rates are corrected
for losses due to pile-up, dead zones on the CCD (i.e.\ bad pixels and bad
columns) and source photons which fell outside the source extraction region.
This correction, which is applied on an event-by-event basis, is achieved by
numerically simulating the PSF for the relevant XRT mode over a radius of 150
pixels, and summing it. It is then summed again, however this time, the
value of any pixel in the simulated PSF which corresponds to a bad pixel in the
data is set to zero before the summation (the lists of bad pixels and the times
for which they were bad were saved in the preparation phase). Furthermore, only
the parts of the PSF which were within the data extraction region are included.
Taking the ratio of the complete PSF to the partial PSF gives the correction
factor. This method is analogous to using exposure maps and the {\sc xrtmkarf}
task, as is done when manually creating light curves. Alternative methods of
using {\sc xrtmkarf} give correction factors which differ by up to 5\%; we
compared our correction factors with these, and found them to lie in the middle
of this distribution. 

In addition to these corrections, we need to ensure that the exposure time is
calculated correctly: mode switching, or bins spanning multiple snapshots, will
result in a bin duration which is much longer than the exposure time. This is
done by using the Good Time Interval (GTI) information from the event lists: if
a bin spans multiple GTIs the dead-times between GTIs are summed, and the result
is subtracted from the bin duration to give the exposure time, which is used to
calculate the count rate. The fractional exposure is defined as the exposure
time divided by the bin duration.

Finally, the data are written to ASCII files. The following information is
saved for each bin:

\begin{itemize}
\item{Time in seconds (with errors). The bin time is defined as the mean photon arrival
time, and the (consequentially asymmetric) errors span the entire time interval
covered by the bin. Time zero is defined as the BAT trigger time. For
non-\Swift\ bursts, the trigger time given in the GCN circular which announced
the GRB is used as time zero.}
\item{Source count rate (and error) in \cps. This is the final count rate, background
subtracted and fully corrected, with a $\pm$1-$\sigma$ error.}
\item{Fractional exposure.}
\item{Background count rate (and error) in \cps. This is the background count rate scaled
to the source region, with a $\pm$1-$\sigma$ error.}
\item{Correction factor applied to correct for to pile-up, dead zones on the CCD, and source
photons falling outside of the source extraction region.}
\item{Measured counts in the source region.}
\item{Measured background counts, scaled to the source region.}
\item{Exposure time}
\item{Detection significance ($\sigma$), before corrections were applied.}
\end{itemize}

If an upper limit is produced, the measured counts and detection significance
columns refer to the data which have been replaced with an upper limit. The
significance of the upper limit is always 3$\sigma$. 

$\sigma$ is always calculated before the corrections are applied, since it is a
measure of how likely it is that the measured counts, not corrected counts, were
caused by a fluctuation in the background level.

\subsubsection{Counts to flux conversion}
\label{sec:flux}

The conversion from count rates (as in our light curves) to flux requires
spectral information. Since automatic spectral fitting is prone to errors (e.g\
due to local minima of the fit statistic), we refrain from doing this.
Furthermore, accurate flux conversion needs to take into account spectral
variation as the flux evolves, which is beyond the scope of this work.

The GCN reports issued by the \Swift\ team contain a mean conversion factor for
a given burst. These tend to be around $5\times10^{-11}$ erg cm$^{-2}$
count$^{-1}$(0.3--10 keV), suggesting such a value could be used as an
approximate conversion. For 10 \Swift\ bursts between GRB~070110 and GRB~070306,
the mean flux conversion is $5.04\tim{-11}$ erg cm$^{-2}$ count$^{-1}$, with a
standard deviation of $2.61\tim{-11}$ erg cm$^{-2}$ count$^{-1}$.

\subsection{Phase \#3 -- Presentation Phase}
\label{sec:present}

\begin{figure}
\resizebox{\hsize}{!}{\includegraphics[width=8.0cm]{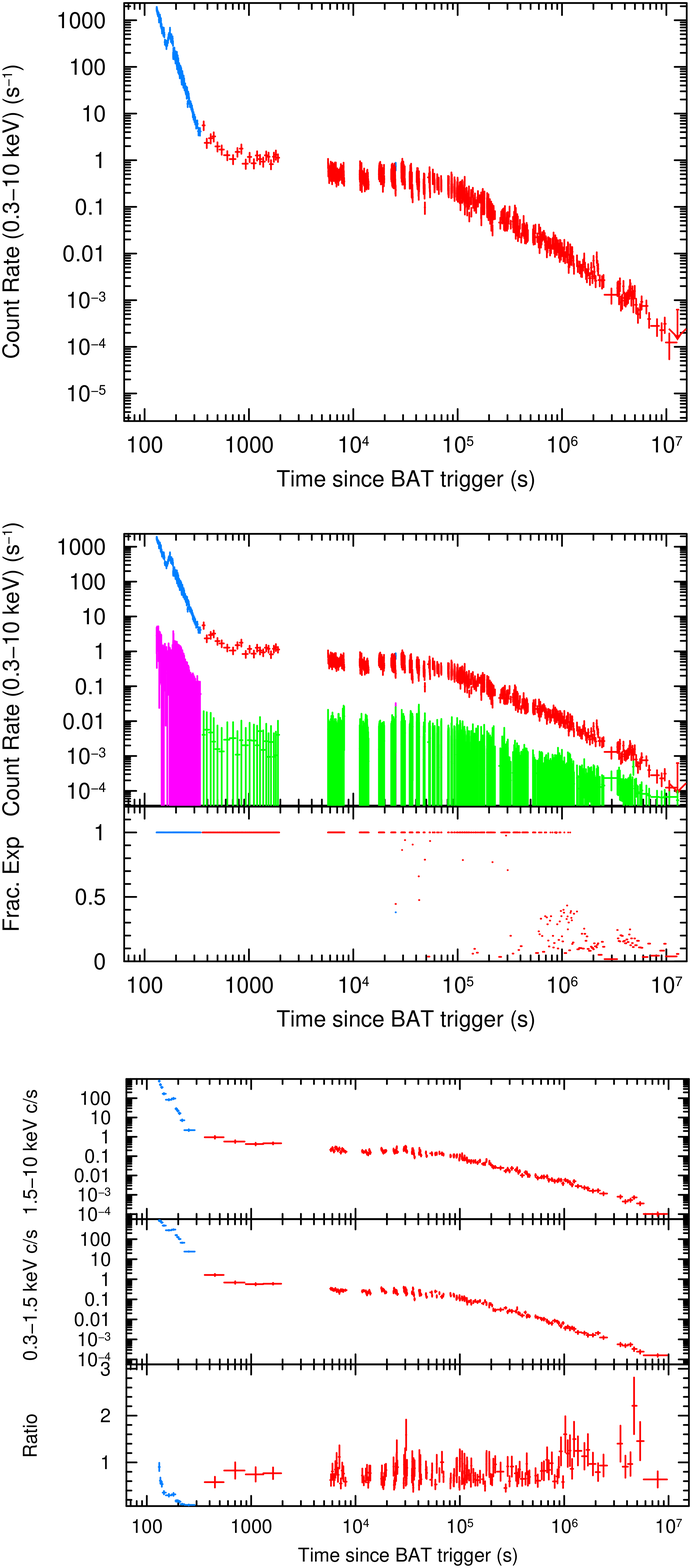}}
%\resizebox{\hsize}{!}{\includegraphics[angle=-90,width=8.0cm]{fig5b.eps}}
%\resizebox{\hsize}{!}{\includegraphics[angle=-90,width=8.0cm]{fig5c.eps}}
\caption{Light curve images for GRB 060729. These data have been discussed by
Grupe \etal(2007).
\newline \emph{Top panel}: Basic light curve.
\newline \emph{Centre panel}: Detailed light curve, with background levels shown below
the light curve, and fractional exposure given in the lower pane.
\newline \emph{Bottom panel}: Hardness ratio. The 3 panes are (top to bottom) Hard: (1.5--10
keV) , Soft (0.3--1.5 keV) and the ratio (Hard/Soft).
}
\label{fig:3curves}
\end{figure}

The final phase parses the output of the production phase to produce
light curves. Three such curves are produced, and Fig.~\ref{fig:3curves} shows an
example of each; count-rates and the time since trigger are plotted
logarithmically. The first is a basic light curve, simply showing count-rate
against time. The second also shows the background level and fractional
exposure. In WT mode when the GRB is bright, the background tends to be
dominated by the $<1$\%\ of the PSF which leaks into the background region, but
because of the high source count rate, this has negligible effects on the
corrected count rates. The PC mode background should generally be approximately
constant. If it shows large variations, the data may be contaminated by enhanced
background linked to the sunlit Earth. Unfortunately, such contamination is
currently unpredictable and varies both spatially and temporally; it is thus
very difficult to correct for manually, and our automated processing does not
currently correct for this. PC mode data points which occur during times of
variable background should thus be treated with caution. We note, however, that
our testing procedure (Section~3) does not show our light curves to be degraded
when bright Earth characteristics become apparent.

The third light curve produced in this phase is energy-resolved. The hard- and
soft-band light curves are shown separately, and the hard/soft ratio makes up the
bottom panel of this plot.

Also created in this phase is a deep PC mode image, using the summed PC event
list created in the preparation phase. This image is split into three energy
bands: 0.3--1.2 keV, 1.2--1.8 keV and 1.8--10 keV. These bands were chosen based
on the spectra of the GRBs seen by \Swift\ to date, to ensure that for a
`typical' burst, there will be approximately equal numbers of counts in each
band.
% defined so that there are equal numbers of counts from the GRB in each
%band. Note that this will be different for each burst. 
These three energy-resolved images are plotted on a logarithmic scale, and
combined (using {\sc ImageMagick} to produce a 3-colour image (with red, green
and blue being the soft, medium and hard bands respectively). This is then
smoothed using {\sc ImageMagick}.

%The decision to define the energy bands for
%each burst independently is not simply cosmetic, it allows some spectral
%information to be conveyed graphically. If the majority of background sources
%are red, for example, the GRB is probably comparatively hard.

Once created, these products are transferred to the online repository.

\begin{figure}
\resizebox{\hsize}{!}{\includegraphics{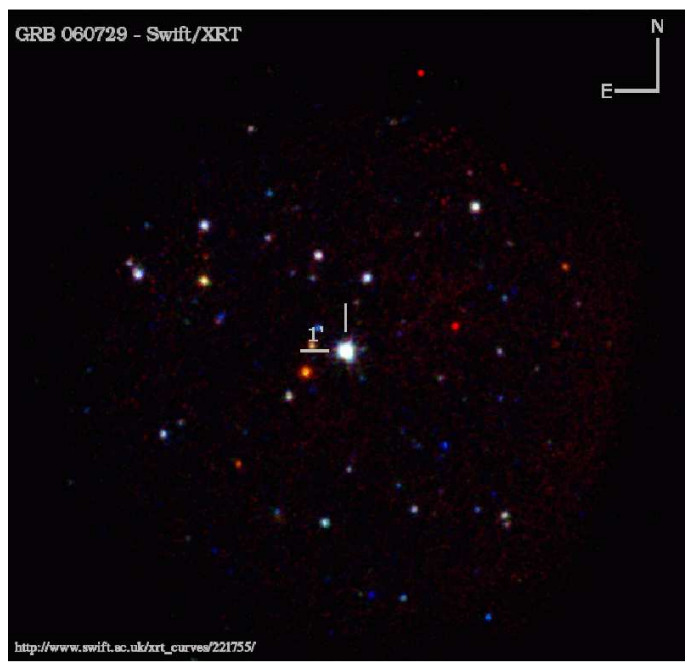}}
\caption{An example three colour image. This is GRB 060729, and the exposure
time is 1.2 Ms.}
\label{fig:deep}
\end{figure}

\subsection{Immediate light curve regeneration}
\label{sec:update}

Our light curve generation is a dynamic process: a light curve is created when
the first XRT data arrive in the UKSSDC archive -- typically 1.5--2 hours after
the burst -- and it is then updated whenever new data have been received and
undergone the pipeline processing. Thus, a light curve should never be more than
\til15 minutes older than the quick-look data. If the GRB is being observed
every orbit, new data can be received as often as every 96 minutes.

The update procedure is identical to that described in Sections 2.1--2.3 above,
except that only the new data are processed and the results appended to the
existing light curve. In the case where the existing light curves ends with an
upper limit, the data from this upper limit are reprocessed with the new data,
hopefully enabling that limit to be replaced with a detection.

\section{Testing procedure}
\label{sec:test}

In order to confirm that our light curves are correct, we used manually created
light curves for every GRB detected by the \Swift\ XRT up to GRB 070306, which
had been produced by one of us (K.L. Page). We broke each light curve up into
phases of constant (power-law) decay, and compared the count-rate and time at
the start and end of each of these phases. We also confirmed that the shape of
the decay was the same in both automatic and manually created light curves. Where
applicable, we also confirmed that the transition between XRT read-out modes
looked the same in both sets of light curves. Once we were satisfied that our
light curves passed this test, we also compared a random sample of 30 GRBs with
those manually created by other members of the \Swift/XRT team and again found
good agreement.

\section{Data availability and usage}
\label{sec:availability}

Our light curve repository is publicly available via the internet, at:

\noindent http://www.swift.ac.uk/xrt\_curves/

\vspace{0.2cm}

Specific light curves can be accessed directly by appending their \Swift\ target
ID to this URL\footnote{The target ID is the trigger number, given in the GCN notices
and circulars, but padded with leading zeroes to be 8 digits long. e.g. GRB
060729 had the trigger number 221755, so its target ID is 00221755}.

While every effort has been made to make this process completely automatic,
there may be cases where the light curve generation fails (e.g. if the source is
too faint to centroid on, or if there are multiple candidates within the BAT
error circle). In this event, a member of the \Swift/XRT team will manually
instigate the creation procedure as soon as possible. For GRBs detected by other
observatories which \Swift\ subsequently observes, the creation procedure
will not be automatically triggered, however the XRT team will trigger
it manually in a timely manner.

These light curves, data and images may be used by anyone. In any publication
which makes use of these data, please cite this paper in the body of your
publication where the light curves are presented. The suggested wording is:

``For details of how these light curves were produced, see Evans \etal(2007).''

Please also include the following paragraph in the Acknowledgements section:

``This work made use of data supplied by the UK Swift Science Data Centre at the
University of Leicester.''

\section{Acknowledgements}

PAE, APB, KLP, LGT and JPO acknowledge PPARC support. LV, JR, PM and DNB are
supported by NASA contract NAS5-00136.

\end{document}